\documentclass[%
 reprint,
superscriptaddress,
 amsmath,amssymb,
 prd,
 altaffilletter,
]{revtex4-2}

\usepackage{graphicx}% Include figure files
\usepackage{dcolumn}% Align table columns on decimal point
\usepackage{bm}% bold math
\usepackage{multirow}
\usepackage{makecell}
\usepackage{xcolor}
\begin{document}

\preprint{APS/123-QED}

\title{Search for Majoron-emitting modes of  $^{136}$Xe double beta decay with the complete EXO-200 dataset}% Force line breaks with \\

\author{S.~Al~Kharusi}\affiliation{Physics Department, McGill University, Montreal, Quebec H3A 2T8, Canada}
\author{G.~Anton}\affiliation{Erlangen Centre for Astroparticle Physics (ECAP), Friedrich-Alexander-University Erlangen-N\"urnberg, Erlangen 91058, Germany}
\author{I.~Badhrees}\altaffiliation{Permanent address: King Abdulaziz City for Science and Technology, Riyadh, Saudi Arabia}\affiliation{Physics Department, Carleton University, Ottawa, Ontario K1S 5B6, Canada}
\author{P.S.~Barbeau}\affiliation{Department of Physics, Duke University, and Triangle Universities Nuclear Laboratory (TUNL), Durham, North Carolina 27708, USA}
\author{D.~Beck}\affiliation{Physics Department, University of Illinois, Urbana-Champaign, Illinois 61801, USA}
\author{V.~Belov}\affiliation{Institute for Theoretical and Experimental Physics named by A.I. Alikhanov of National Research Centre ``Kurchatov Institute'', Moscow 117218, Russia}
\author{T.~Bhatta}\altaffiliation{Present address: Department of Physics and Astronomy, University of Kentucky, Lexington, Kentucky 40506, USA}\affiliation{Department of Physics, University of South Dakota, Vermillion, South Dakota 57069, USA}
\author{M.~Breidenbach}\affiliation{SLAC National Accelerator Laboratory, Menlo Park, California 94025, USA}
\author{T.~Brunner}\affiliation{Physics Department, McGill University, Montreal, Quebec H3A 2T8, Canada}\affiliation{TRIUMF, Vancouver, British Columbia V6T 2A3, Canada}
\author{G.F.~Cao}\affiliation{Institute of High Energy Physics, Beijing 100049, China}
% ...

\date{\today}% It is always \today, today,
             %  but any date may be explicitly specified

\begin{abstract}
A search for Majoron-emitting modes of the neutrinoless double-beta decay of $^{136}$Xe is performed with the full EXO-200 dataset. This dataset consists of a total $^{136}$Xe exposure of 234.1 kg$\cdot$yr, and includes data with detector upgrades that have improved the energy threshold relative to previous searches. A lower limit of T$_{1/2}^{\rm{^{136}Xe}}>$4.3$\cdot$10$^{24}$ yr at 90\% C.L. on the half-life of the spectral index $n=1$ Majoron decay was obtained, a factor of 3.6 more stringent than the previous limit from EXO-200 and a factor of 1.6 more stringent than the previous best limit from KamLAND-Zen. This limit corresponds to a constraint on the Majoron-neutrino coupling constant of $|\langle g_{ee}^{M}\rangle|$$<(0.4$-$0.9)\cdot10^{-5}$. The lower threshold and the additional data taken resulted in a factor 8.4 improvement for the $n=7$ mode compared to the previous EXO-200 search. This search provides the most stringent limits to-date on the Majoron-emitting decays of $^{136}$Xe with spectral indices $n=1,2,3,$ and 7.

\end{abstract}
%\keywords{Suggested keywords}%Use showkeys class option if keyword
                              %display desired
\maketitle

\section{\label{sec:intro}Introduction}
Double-beta ($\beta\beta$) decay is a rare weak transition between two nuclei with the same mass number $A$ and nuclear charges $Z$ that differ by two units. The process is only observable if single-beta ($\beta$) decay is highly suppressed or forbidden by energy conservation. Decays in which two neutrinos are emitted ($2\nu\beta\beta$) are an allowed process in the Standard Model, and have been observed in a number of nuclides~\cite{particle2020review} including $^{136}$Xe with a half-life of $T^{2\nu\beta\beta}$=[$2.165 \pm 0.016$~(stat)~$\pm 0.059$~(syst)]$\times 10^{21}$~yr~\cite{PhysRevC.89.015502}. However, if neutrinos are massive Majorana fermions, $\beta\beta$ decays can also proceed without emission of neutrinos, violating lepton number conservation~\cite{PhysRevD.25.2951}. The simplest of such modes, the 0$\nu\beta\beta$ decay with the emission of two electrons and nothing else, is a subject of an intense experimental search~\cite{Dolinski_review:2019nrj}. The most recent measurements have set stringent lower limits on the half-life for 0$\nu\beta\beta$ decay of several isotopes, including $^{136}$Xe~(EXO-200~\cite{PhysRevLett.123.161802} and KamLAND-Zen~\cite{KamLAND-Zen:2016pfg}), $^{76}$Ge~(GERDA~\cite{GERDA:2020xhi}), and $^{130}$Te~(CUORE~\cite{CUORE:2021gpk}).
\par
\begin{figure}[tbp]
\begin{center}
\includegraphics[width=0.5\textwidth]{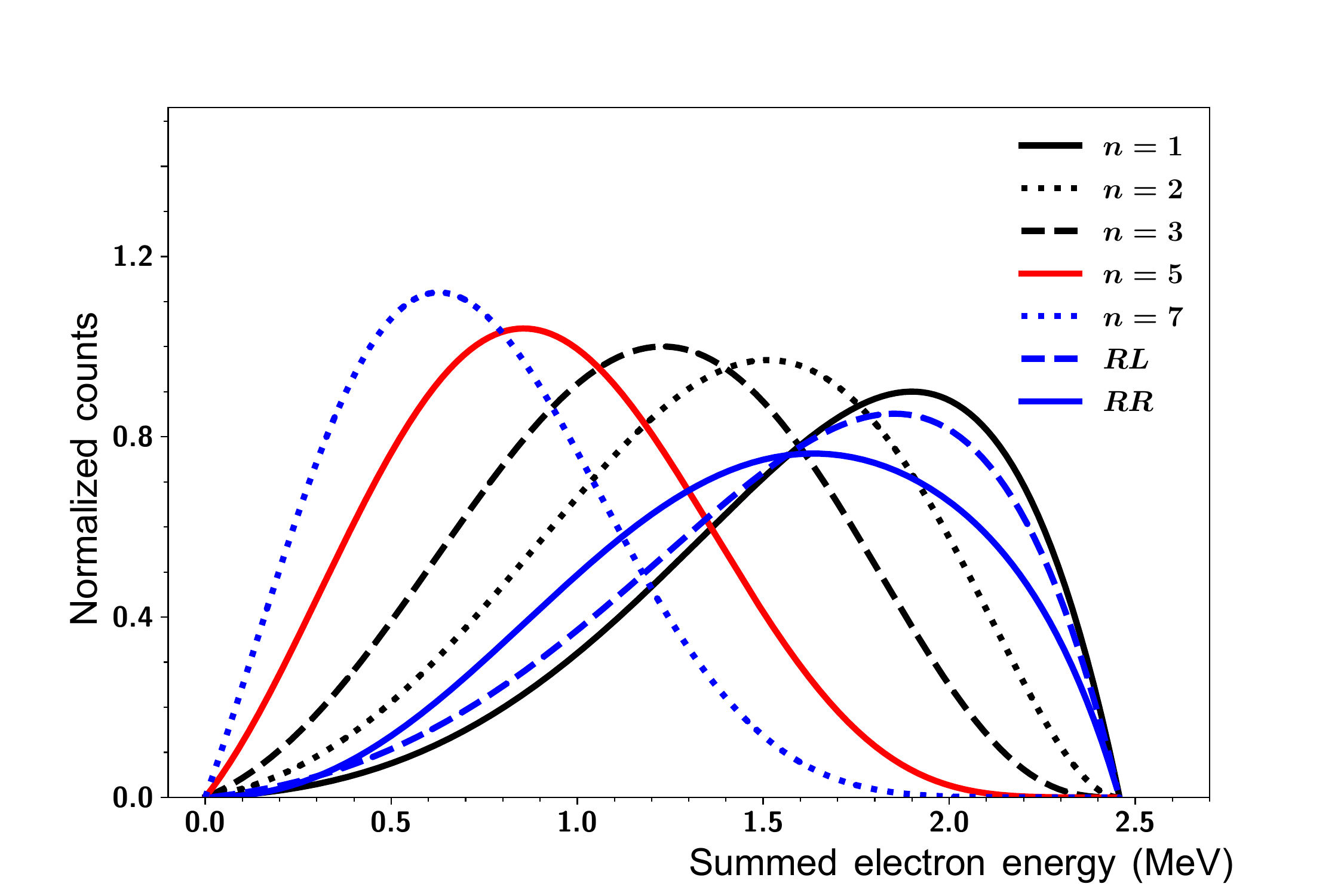}
\caption{Energy spectra for the $n=1,2,3,$ and 7 Majoron decay modes of $^{136}$Xe. Also shown are the $2\nu\beta\beta$ decay spectrum ($n=5$), along with spectra for a Majoron-like scalar particle, $\phi$, in a novel model of $0\nu\beta\beta$ decay where it couples via an effective seven-dimensional operator with a right-handed lepton current and right-handed ($\epsilon^{\phi}_{RR}$) and left-handed ($\epsilon^{\phi}_{RL}$) quark current as proposed in~\cite{PhysRevLett.122.181801}.}
\label{fig:majoron_spectra}
\end{center}
\end{figure}

 In this paper we present results of a search for neutrinoless $\beta\beta$ decay modes of $^{136}$Xe in which one or two additional bosons, denoted as $\chi_0$ here, are emitted together with the electrons, i.e.:
\begin{equation}
(A,Z)\rightarrow (A,Z+2)+2e^-+\chi_0
\end{equation}
or
\begin{equation}
(A,Z)\rightarrow (A,Z+2)+2e^-+2\chi_0
\end{equation}
Any bosons emitted in the 0$\nu\beta\beta\chi_0$ (0$\nu\beta\beta\chi_0\chi_0$) modes are usually referred to as ``Majorons''. Majorons were originally proposed as a Goldstone boson associated with spontaneous lepton number symmetry breaking~\cite{Chikashige:1980ui,Gelmini:1980re,Georgi:1981pg}. Majorons are possible dark matter candidates~\cite{Lattanzi:2013uza}, and may be involved in cosmological and astrophysical processes~\cite{Dolgov:2004jw,PhysRevD.84.105040}. Precise measurement of the width of the $Z$ boson decay to invisible channels~\cite{2006257} has disfavored the original Majoron models. However, other analogous models have been proposed, free of this constraint, in which Majorons more generally refer to massless or light bosons that might be neither Goldstone bosons, nor be required to carry a lepton charge~\cite{barabash2015double}. The spectral index $n$ is used to characterize different Majoron-emitting modes that are experimentally recognizable by the shape of the sum electron spectrum~\cite{BAMERT199525,HIRSCH19968}. A novel model of $0\nu\beta\beta$ decay with a emission of a light Majoron-like scalar particle $\phi$ was also proposed in~\cite{PhysRevLett.122.181801}, where the Majoron-like particle couples via an effective seven-dimensional operator with a right-handed lepton current, and with right-handed ($\epsilon^{\phi}_{RR}$) or left-handed ($\epsilon^{\phi}_{RL}$) quark current. The normalized spectra for various Majoron-emitting decay modes of $^{136}$Xe are shown in Fig.~\ref{fig:majoron_spectra}. The calculation of the spectra uses the Fermi function suggested in~\cite{schenter1983simple} that fully includes the nuclear finite size and electron screening~\cite{Albert:2014fya} and evaluates the value of the Fermi function at the nuclear radius $R$ as recommended in~\cite{Kotila:2012zza}. The single-state dominance model of $2\nu\beta\beta$ is used, while the higher-state dominance models that yield slightly different spectral shapes~\cite{Moreno:2008dz,PhysRevC.87.014315} are not considered in the analysis.  

\par Recent sensitive searches for Majoron-emitting $\beta\beta$ decays have been carried out in $^{76}$Ge~(GERDA~\cite{Agostini:2015nwa}), $^{130}$Te~(CUORE~\cite{davis2020search}), and $^{136}$Xe~(KamLAND-Zen~\cite{KamLAND-Zen:2016pfg} and EXO-200~\cite{Albert:2014fya}). EXO-200 reported a lower limit of $T_{1/2}^{\rm{^{136}Xe}}>1.2\times 10^{24}$ yr at 90\% C.L. on the half-life of the spectral index $n=1$ Majoron decay mode based on 100~kg$\cdot$yr exposure of $^{136}$Xe~\cite{Albert:2014fya}, compared with a lower limit of $2.6\times10^{24}$~yr at 90\% C.L. reported by KamLAND-Zen~\cite{KamLAND-Zen:2016pfg}. Following that analysis, several detector upgrades were made to EXO-200 permitting a lower analysis threshold, and additional data were acquired in ``Phase-II'' of EXO-200 operations from May 2016 to December 2018 utilizing these technical improvements. The total Phase-II exposure collected was similar to that of the first run~(``Phase-I'', September 2011 to February 2014) from which the previous searches for Majoron-emitting modes were reported. This paper reports a search for Majoron-emitting modes of $\beta\beta$ decay using the full EXO-200 dataset that totals 1181.3 days of live time after data quality cuts described in~\cite{Albert:2013gpz}, corresponding to a 134\% increase in $^{136}$Xe exposure relative to the previous search~\cite{Albert:2014fya}.

\section{\label{sec:detector}Detector description, Data and Monte Carlo Simulation}
The EXO-200 detector consisted of a cylindrical liquid xenon (LXe) time projection chamber (TPC) filled with LXe enriched to 80.6\% in $^{136}$Xe. A cathode split the TPC into two drift regions, each with a radius of $\sim$18~cm and a drift length of $\sim$20~cm. The TPC was enclosed by a radiopure thin-walled copper vessel. The electric field in the drift region was raised from 380~V$/$cm in Phase-I to 567~V$/$cm in Phase-II to improve the energy resolution. The ionization produced from
interactions in each drift region was read out after being drifted to crossed-wire planes at each anode, inducing signals on the front-most wire plane (V-wires), after which it was collected by the second wire plane (U-wires). The scintillation light was collected by arrays of large area avalanche photo-diodes (LAAPDs)~\cite{NEILSON200968} located behind the wire planes. A more detailed description of the detector can be found in~\cite{Auger_2012,Ackerman:2021ijn}.
\par The detector was located inside a clean room at the Waste Isolation Pilot Plant (WIPP) in Carlsbad, New Mexico, under an overburden of 1623$^{+22}_{-21}$ meters water equivalent~\cite{Albert_2016}. An active muon veto system consisting of plastic scintillator panels surrounding the clean room on four sides allowed prompt identification of $>$94\% of the cosmic ray muons passing through the setup, and allowed rejection of cosmogenic backgrounds~\cite{Albert_2016}. 
\par Each TPC event is reconstructed by grouping charge and light signals into individual energy deposits. Ionization signals measured by the two wire planes provide information about coordinates $x$ and $y$ perpendicular to the drift field. The time difference between the light signal and the associated charge signal, together with the measured electron drift velocity~\cite{PhysRevC.95.025502}, provides the $z$ position. Events with a single reconstructed charge deposit are referred to as ``single site'' (SS), and include most $\beta$ or $\beta\beta$ decays with characteristic spatial extent of~2--3~mm. Events with multiple reconstructed deposits are referred to as ``multisite'' (MS), and arise mostly from multiple interactions of MeV-energy $\gamma$-rays. Additionally, internally generated $\beta\beta$-like events in the fiducial volume (FV) are uniformly distributed in the LXe, compared to the spatial distribution of background events arising from $\gamma$-rays entering the TPC, which tend to be concentrated nearer to the vessel walls. This difference is captured in the analysis by the standoff-distance (SD) variable, defined as the shortest distance between any reconstructed charge deposit and the closest material surface excluding the cathode. The total energy of an event is determined by combining the charge and scintillation signals. This combination achieves better energy resolution than possible from each individual channel alone due to the anticorrelation between them~\cite{PhysRevB.68.054201,Anton:2019hnw}. 
\par The detector response to $\beta\beta$ decays and background interactions is modeled by a detailed Monte Carlo (MC) simulation based on GEANT4~\cite{1610988}. Radioactive $\gamma$ sources are deployed at several positions near the TPC to characterize the detector response and validate the MC simulation. The scintillation and ionization yields were determined with $\gamma$ interactions from calibration sources over a range of electric fields~\cite{Anton:2019hnw}. The energy scale and resolution are simultaneously determined by fitting the expected energy spectra generated by MC to the corresponding calibration $\gamma$ sources. The absolute $\beta\beta$ energy scale is found to be consistent with the calibration $\gamma$ sources at the sub-percent level~\cite{PhysRevLett.123.161802,Anton:2019hnw}.
\par The dataset and event selection criteria used in this work is the same as in the search for $0\nu\beta\beta$ decay~\cite{PhysRevLett.123.161802}, except that a reduced energy threshold is used here. The $^{136}$Xe exposure of the entire dataset after data quality cuts and accounting for live time loss due to vetoing events coincident with the muon detector is 234.1~kg$\cdot$yr, or 1727.5~mol$\cdot$yr, with 117.4 (116.7)~kg$\cdot$yr in Phase-I (Phase-II).

\begin{figure*}[t]
\begin{center}
         \includegraphics[width=0.45\textwidth]{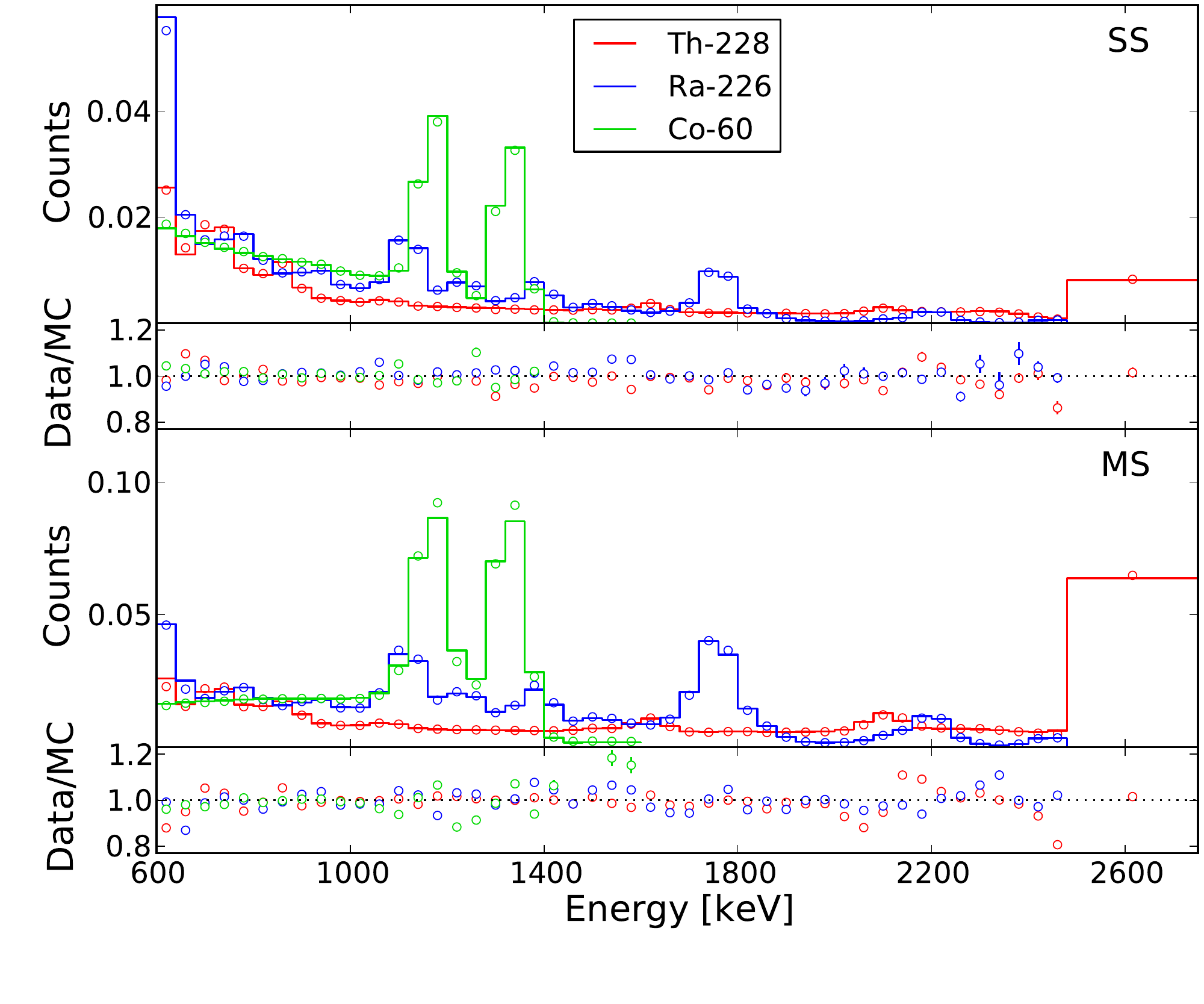}
         \includegraphics[width=0.45\textwidth]{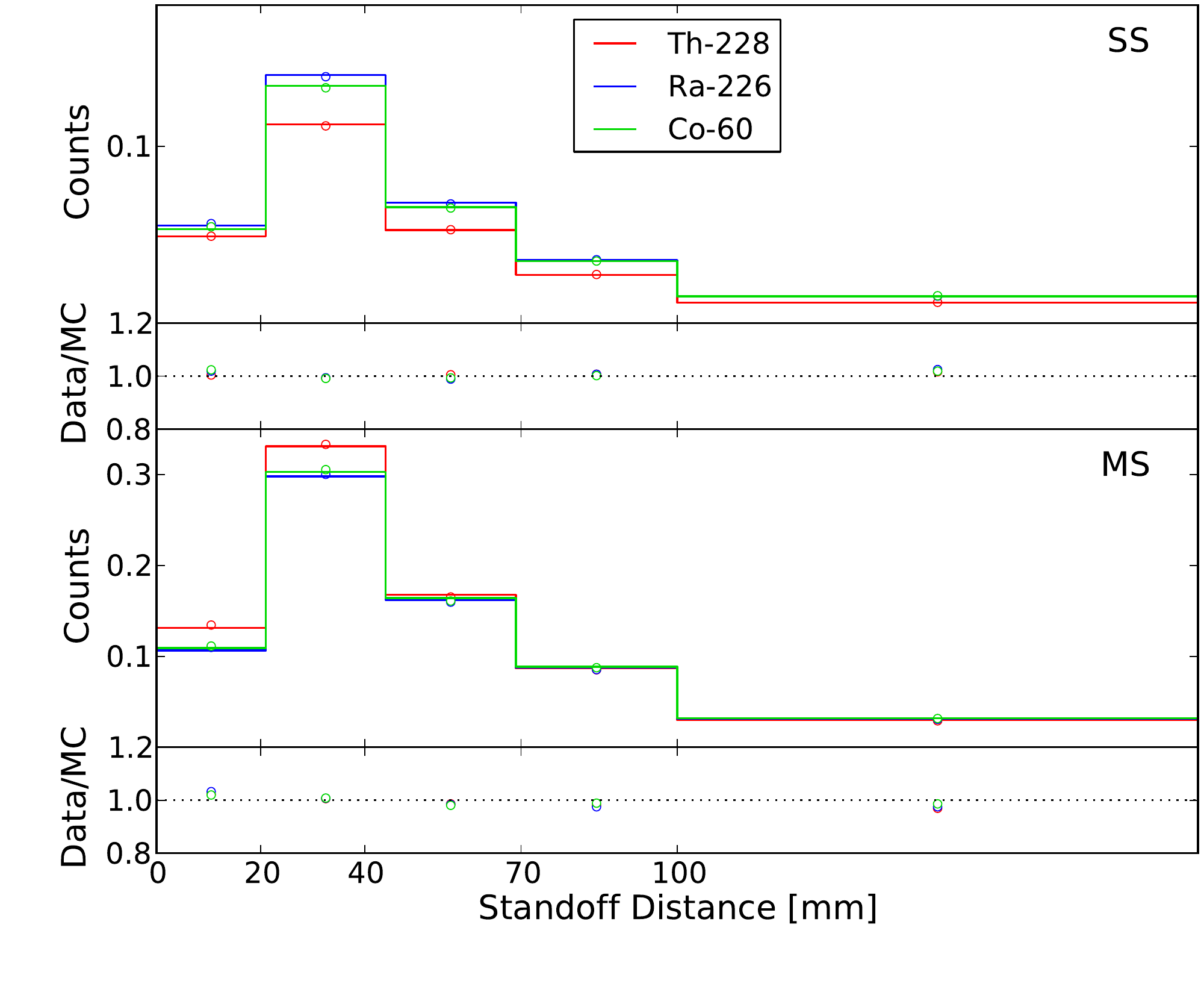}
\caption{Comparison between radioactive source data (circles) and MC (lines) for energy (left) and SD (right) distributions for SS (top) and MS (bottom) events from calibration sources positioned near the cathode in Phase-II. The event count in the last bin of energy distributions contains overflow events outside the plotted range and is multiplied by a factor of 0.2 for visibility.}
\label{fig:agreements}
\end{center}
\end{figure*}
\section{\label{sec:data}Analysis Procedure}
The energy threshold used in this work is lowered from the 1000~keV threshold used in the most recent $0\nu\beta\beta$ search by EXO-200~\cite{PhysRevLett.123.161802} to 600~keV, to improve sensitivity to Majoron-emitting modes at low energy. As shown in Fig.~\ref{fig:agreements}, the energy measurement shows good spectral agreement between the data and the simulation for SS and MS events above 600~keV using $^{228}$Th, $^{226}$Ra, and $^{60}$Co calibration sources. The SD is also observed to have good agreement between the data and the simulation at energies above 600~keV, as shown in Fig.~\ref{fig:agreements}.

\par Probability density functions (PDFs) for signal and background components are generated using the Monte Carlo~(MC) simulation. The PDFs are functions of two observables: energy and SD. Studies were performed to estimate the sensitivity improvement possible with additional multivariate discriminators similar to those used in previous $0\nu\beta\beta$ searches~\cite{PhysRevLett.123.161802,Albert:2017owj}, but the minimal set of variables (energy and SD) was chosen as it had comparable sensitivity while minimizing systematic errors at low energy. The systematic errors in the analysis are described in detail in Sec.~\ref{sec:systematicerror}. To avoid any possible bias in analysis criteria, all selection cuts and the choice of fitting variables were determined from MC-based sensitivity studies alone prior to performing any fits to the data itself. The components of the overall fit model are similar to that in~\cite{PhysRevLett.123.161802} with the $0\nu\beta\beta$ signal replaced by a Majoron-emitting decay. Since this work used a reduced energy threshold, two components are added to the background model:
\begin{itemize}
    \item $^{85}$Kr dissolved in the LXe that produces $\beta$ decays with an end point energy of 687.0~keV~\cite{nudat2}
    \item $^{137}$Cs in the materials near the LXe, which emits $\gamma$-rays with energy of 661.7~keV~\cite{nudat2}
\end{itemize}
The simulation of $^{85}$Kr includes the two $\beta$ decay modes with branching ratios of 99.56\% and 0.434\% to the ground and excited states of $^{85}$Rb followed by the release of a 514.0~keV $\gamma$-ray, respectively~\cite{nudat2}. A shape correction accounting for the forbidden nature of the first unique $\beta$ decay was calculated using the method described in~\cite{konopinski1966theory} to be between -15\% and 80\% depending on its energy. This correction was applied as an event weighting in the MC simulation.
\par The PDF model is parametrized by the event counts and SS fractions ($f_{SS}$ = SS/(SS+MS)) of the individual components, as well as two normalization parameters to account for the effects of systematic errors~\cite{Albert:2017feo}. The search was performed using a maximum-likelihood (ML) function to fit simultaneously both SS and MS events with their corresponding PDFs generated by MC, in a similar approach as ~\cite{PhysRevLett.123.161802}. Systematic errors, described in Sec.~\ref{sec:systematicerror} are included in the ML fit as nuisance parameters constrained by normal distributions. The median 90\% CL sensitivity is estimated using toy datasets generated from the PDFs of the background model. An energy threshold of 600~keV is used in the fit, which provides near optimal sensitivity for all Majoron-emitting modes considered here. The reduced energy threshold relative to previous analyses~\cite{Albert:2014fya,PhysRevLett.123.161802} results in higher signal detection efficiencies, especially for the $n=7$ mode, which has a peak around 628~keV in the energy spectrum. Lower energy thresholds do not further improve sensitivity because the increase in signal efficiency is outweighed by the presence of backgrounds, including $^{85}$Kr.

\section{\label{sec:systematicerror}Systematic Errors}
Systematic errors were accounted for by the same technique as described in~\cite{Albert:2017feo, PhysRevLett.123.161802}. The five Gaussian constraints added to the ML fit, which are used to propagate systematic errors into the results, correspond to:
\begin{itemize}
    \item uncertainty in the activity of radon in the LXe as determined in stand-alone studies via measurement of time correlated $^{214}$Bi-$^{214}$Po decays~\cite{PhysRevC.92.015503};
    \item uncertainty in the relative fractions of neutron capture-related PDF components generated by dedicated simulations~\cite{Albert_2016};
\item uncertainty in SS fractions obtained by comparisons between calibration data and MC;
\item uncertainty in the overall event detection efficiency, referred to as \textit{normalization}, caused by event reconstruction and selection efficiencies;
\item uncertainty in the signal detection efficiency, referred to as \textit{signal-specific normalization}, caused by discrepancies in the shape distributions between data
and MC and background model uncertainties.
\end{itemize}
\begin{figure}[t]
\begin{center}
\includegraphics[width=0.4\textwidth]{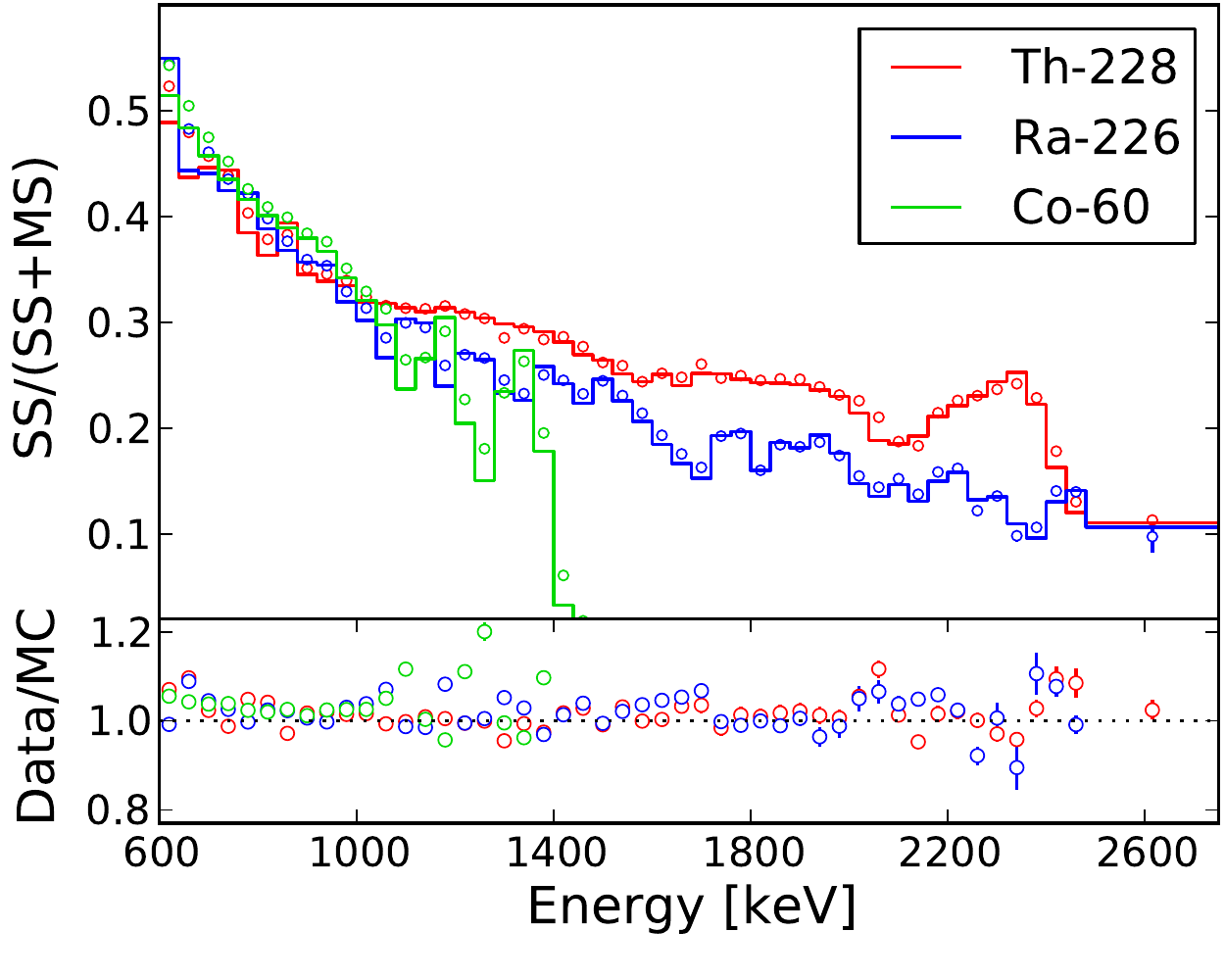}
\caption{SS fractions for MC (lines) and data (circles) in Phase-II using calibration sources positioned near the cathode.}
\label{fig:ssfraction}
\end{center}
\end{figure}
The first two uncertainties were constrained by relative errors of 10\% and 20\%, respectively, as evaluated in~\cite{exo2014prc}. The uncertainty in the SS fractions is determined by comparisons between the data and MC for calibration sources, as shown in Fig.~\ref{fig:ssfraction}. Taking into account different calibration sources at various positions, these systematics are evaluated to be 3.7\% (3.6\%) for Phase-I (Phase-II), averaged over the energy range considered here. The uncertainty on the overall efficiency was evaluated to be 3.1\% (3.0\%) for Phase-I (Phase-II), and differs slightly from that estimated in~\cite{PhysRevLett.123.161802} after accounting for the larger energy range considered here. The top four uncertainties are presented in Table~\ref{tab:sys_common}.
\par Discrepancies in the shape distributions between data and MC are propagated into the signal rate through a normalization parameter that only scales the coefficient of the signal PDFs. This signal-specific normalization parameter is constrained to unity within the errors arising from spectral shape agreement and background model uncertainties as described below. To estimate the effect of spectral shape errors, the ratio between data and MC of the projections onto energy and SD (Fig.~\ref{fig:agreements}) were used to re-weight all PDF components by the observed differences (referred to below as \textit{un-skewing}). $^{60}$Co and $^{238}$U-related PDFs were weighted by ratios from $^{60}$Co and $^{226}$Ra calibration sources respectively, while the other $\gamma$-like PDFs were weighted by ratios from the $^{228}$Th source that has the most data. Toy datasets were generated from these un-skewed background PDFs, along with a given number of signal events. These toy datasets were then fit with the standard background and signal PDFs used in the primary analysis. The average difference between the true number of signal events added to the toy datasets and the best-fit signal counts is used to quantify the impact of the spectral discrepancy. To evaluate the uncertainty associated with the background model, decays of Th, U, and Co were simulated in different locations than that in the default model. For example, all far $^{238}$U are represented by the decays in the air gap between the cryostat and the lead shielding in the background model. To evaluate the errors introduced by this approximation, $^{238}$U simulated in the cryostats is used to represent all $^{238}$U from remote locations. This is taken to represent an extreme deviation from the more realistic case used in this analysis. Toy datasets generated with the default model along with a given number of signal events were fitted with this alternative model, and the resulting difference between the true number of signal events added to the toy datasets and the best-fit signal counts is taken as the systematic error of the background model. The background model error also includes the effects of perturbations to the 2$\nu\beta\beta$ spectrum due to corrections to the Fermi function arising from the finite nuclear size and electron screening effects~\cite{schenter1983simple,Kotila:2012zza}. The $2\nu\beta\beta$ PDF integrals differed by 1.5\% in the case of a differing Fermi function. The signal-specific normalization error is a function of the number of signal counts to account for possible existence of signal events, that is presented in Table~\ref{tab:sys_mode}.

\begin{table}[t]
\caption{Summary of the top four systematic errors added to the searches for Majoron-emitting decays of $^{136}$Xe that are common for different modes in Phase-I and Phase-II.}
\begin{center}
\begin{tabular}{c c c }
\hline
\hline
Constraints & Phase-I & Phase-II\\
\hline
Radon in LXe&10\%&10\%\\
Neutron-capture PDF fractions &20\%&20\%\\
SS fractions&3.7\%& 3.6\% \\
Normalization&3.1\%& 3.0\%\\
\hline
\hline
\end{tabular}
\end{center}
\label{tab:sys_common}
\end{table}%

\begin{table}[t]
\caption{Summary of the signal-specific normalization relative error that is calculated by $\sigma/N=\sqrt{(a\cdot N)^2+b^2}/N$, where $N$ is the number of signal counts, in Phase-I and Phase-II.}
\begin{center}
\begin{tabular}{c c c c c c c c}
\hline
\hline
 Decay Mode& & $n=1$& $n=2$ & $n=3$ & $n=7$& RR & RL\\
\hline

\makecell{Phase-I}& \makecell{a\\b}&\makecell{0.11\\138} & \makecell{0.04\\201}& \makecell{0.08\\435} & \makecell{0.37\\2040} &\makecell{0.25\\62}&\makecell{0.17\\66}\\
\makecell{Phase-II}&\makecell{a\\b}&\makecell{0.19\\48} & \makecell{0.12\\123}& \makecell{0.18\\143} & \makecell{0.02\\555} &\makecell{0.27\\37}&\makecell{0.22\\47}\\
\hline
\hline
\end{tabular}
\end{center}
\label{tab:sys_mode}
\end{table}%

\section{\label{sec:result}Results and conclusion}
\begin{figure*}[t]
\begin{center}
\includegraphics[width=0.45\textwidth]{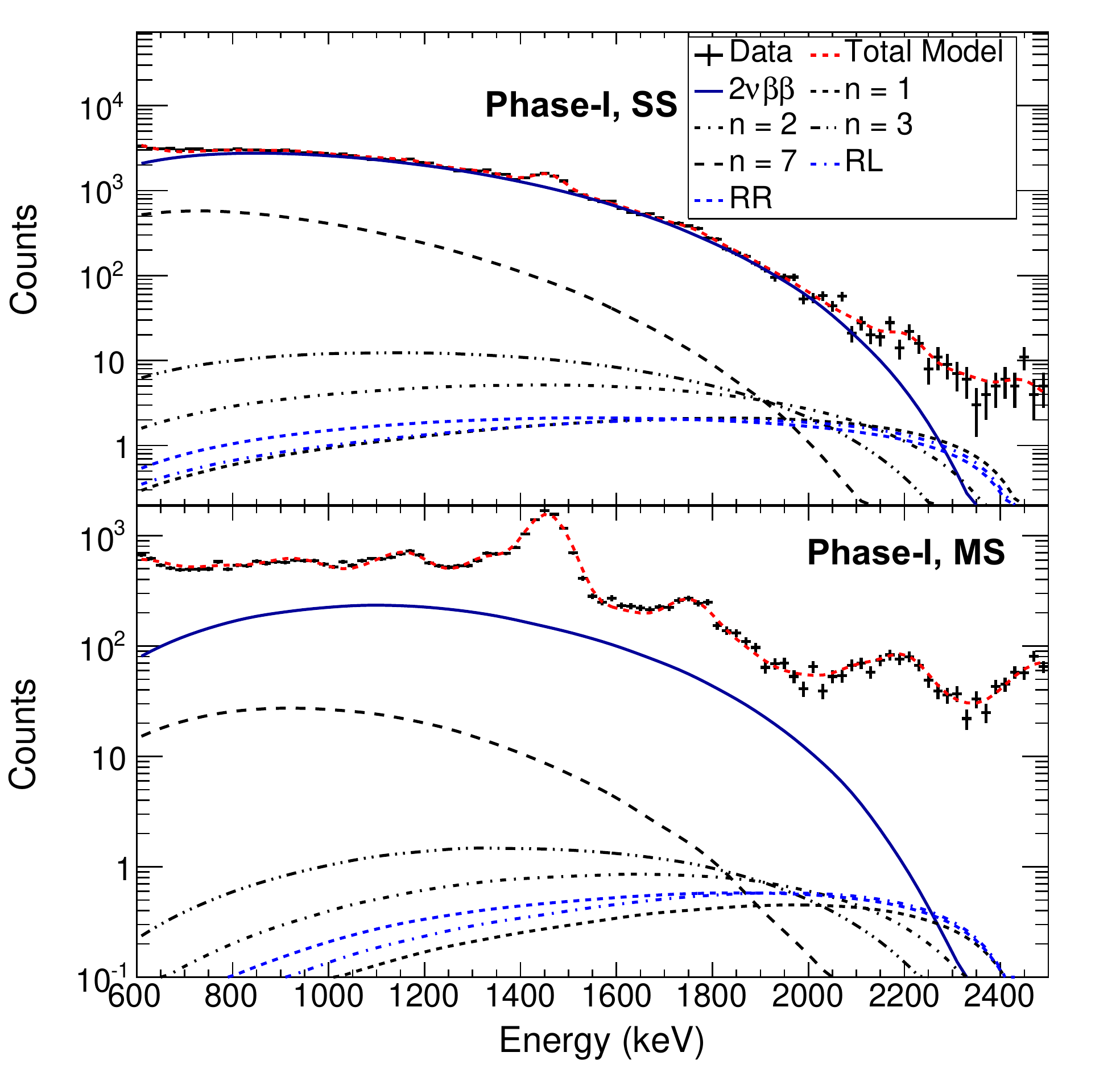}
\includegraphics[width=0.45\textwidth]{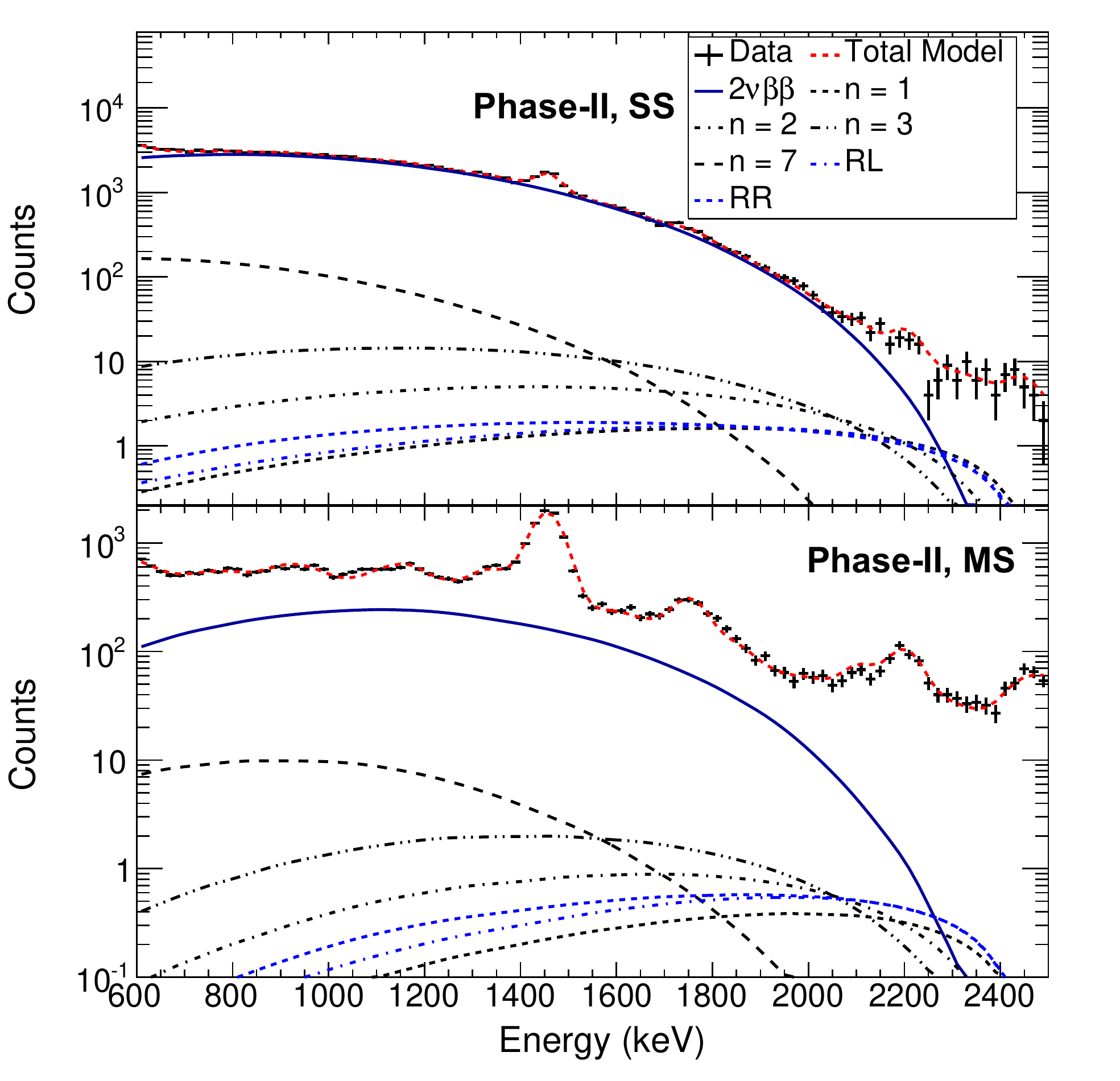}
\caption{SS (top) and MS (bottom) datasets and the best-fit models for the case of the $n=1$ Majoron fit in Phase-I (left) and Phase-II (right). SS energy is predominantly populated by $\beta$-like events. The upper 90\% C.L. limits on the number of decays for each of the six Majoron emitting modes are calculated separately, but here are plotted all at once as an illustration.}
\label{fig:fit_p2}
\end{center}
\end{figure*}
To determine confidence intervals on each of the Majoron decay modes considered here, the datasets from Phase-I and Phase-II are fit independently with efficiency and livetime from each phase taken into account. The observed data, and an example of the best fit spectrum for the $n=1$ Majoron mode, are shown in Fig.~\ref{fig:fit_p2}. No statistically significant evidence for Majoron-emitting $0\nu\beta\beta$ decays is observed for any mode considered. The fits are initially performed individually for each Majoron-emitting decay mode separately for Phase-I and Phase-II, and the combined limits are determined after summing the profile likelihood obtained from each dataset. While the fits are performed for each mode independently, Fig.~\ref{fig:fit_p2} also overlays the corresponding spectra from all fits at the 90\% C.L. upper limits on the number of decays for each mode. The lower limits on the Majoron-emitting $0\nu\beta\beta$ half-lives are summarized in Table~\ref{tab:limits}. EXO-200 has better detection efficiency at low energies and improved spectral agreement between data and MC simulation in Phase-II with upgraded electronics, which result in more stringent lower limits on most Majoron-emitting $0\nu\beta\beta$ half-lives than in Phase-I. The improvement is particularly significant for $n=7$ mode that has a spectrum peaking at lower energy than the other modes.

\begin{table*}[ht]
\caption{90\% C.L. limits on half-lives and coupling constants for different Majoron decay models, and comparison to the previous EXO-200 results~\cite{Albert:2014fya}.}
\begin{center}
\begin{tabular}{c c c c c c }
\hline
\hline
Decay Mode &  Phase-I, yr & Phase-II, yr & Combined, yr &  $|\langle g_{ee}^M\rangle|$ &  EXO-200 (2014), yr  \\
\hline
\rule{0pt}{3ex}    $0\nu\beta\beta\chi_0$ $n=1$&$>$2.3$\times$10$^{24}$ & $>$3.0$\times$10$^{24}$ & $>$4.3$\times$10$^{24}$ & $<(0.4$-$0.9)\cdot10^{-5}$ &$>$1.2$\times$10$^{24}$\\
$0\nu\beta\beta\chi_0$ $n=2$&$>$9.7$\times$10$^{23}$ & $>$9.8$\times$10$^{23}$ & $>$1.5$\times$10$^{24}$ & - & $>$2.5$\times$10$^{23}$\\\
$0\nu\beta\beta\chi_0$ $n=3$&$>$4.6$\times$10$^{23}$ & $>$3.8$\times$10$^{23}$ & $>$6.3$\times$10$^{23}$ & $<0.01$ & $>$2.7$\times$10$^{22}$\\\
$0\nu\beta\beta\chi_0\chi_0$ $n=3$&$>$4.6$\times$10$^{23}$ & $>$3.8$\times$10$^{23}$ & $>$6.3$\times$10$^{23}$ & $<(0.3$-$2.5)$  & $>$2.7$\times$10$^{22}$\\\
$0\nu\beta\beta\chi_0\chi_0$ $n=7$&$>$1.6$\times$10$^{22}$ & $>$6.0$\times$10$^{22}$ & $>$5.1$\times$10$^{22}$ & $<(0.3$-$2.8)$ & $>$6.1$\times$10$^{21}$\\\
	RR&$>$2.0$\times$10$^{24}$ & $>$2.2$\times$10$^{24}$ & $>$3.7$\times$10$^{24}$ & - & -\\
RL&$>$2.3$\times$10$^{24}$ & $>$2.7$\times$10$^{24}$ & $>$4.1$\times$10$^{24}$ & - & -\\
\hline
\hline
\end{tabular}
\end{center}
\label{tab:limits}
\end{table*}%
\par The lower limit on the Majoron-emitting $0\nu\beta\beta$ half-lives for the models with $n= $1, 3, and 7 can be translated into limits on the effective neutrino-Majoron coupling constants $\langle g^M_{ee}\rangle$ using:
\begin{equation}
\frac{1}{T_{1/2}}=|\langle g^M_{ee}\rangle|^{m}\cdot|M^{\prime}|^2\cdot G^{0\nu M}(Z,E_0),
\end{equation}
where $M^{\prime}$=$M(\frac{g_A}{1.25})^2$, $M$ is the nuclear matrix element, $g_A$ is the axial coupling constant, $m=2$ (4) for the emission of one (two) Majorons, and $G^{0\nu M}(Z,E_0)$ is the unnormalized phase space integral that depends on the model type and fundamental constants~\cite{Albert:2014fya}. Table~\ref{tab:limits} summarizes the 90\% C.L. upper limits on the effective Majoron-neutrino coupling constants. The phase space factors are taken from~\cite{Albert:2014fya}, while the matrix elements are taken from~\cite{Menendez:2008jp,Simkovic:2009pp} for the Majoron decay mode with $n=1$, and from~\cite{Hirsch:1995in} for other modes. The spread in the limits on the coupling constants is due to ambiguity in the matrix elements. This is the most stringent limit on $\langle g^M_{ee}\rangle$ for the $n=1$ Majoron among all $\beta\beta$ decay nuclei~\cite{Gando:2012pj,NEMO-3:2013pwo,Agostini:2015nwa,davis2020search}. The previous best limit from a laboratory experiment comes from KamLAND-Zen, which reported a limit of $\langle g^M_{ee}\rangle <$(0.8-1.6)$\cdot 10^{-5}$~\cite{Gando:2012pj}. The phase-space integral for the $n=1$ Majoron used by KamLAND-Zen is about a factor two smaller than the most up to date value used here~\cite{Albert:2014fya}. KamLAND-Zen’s half-life limit would translate to a limit on the coupling constant of $\langle g^M_{ee}\rangle <$(0.6-1.2)$\cdot 10^{-5}$ with the same phase space factor, and our new limit corresponds to a factor of 1.3 improvement over this previous result. Our limit on the coupling constant of $\langle g^M_{ee}\rangle$ is two orders of magnitude more stringent than the recent result obtained in the measurement of pion decays~\cite{PIENU:2021clt}.
\par In conclusion, we report results from a search for Majoron-emitting double-beta decay modes of $^{136}$Xe with the complete EXO-200 dataset. No statistically significant evidence for this process is found, and we obtain limits on half-lives and effective coupling constants of Majoron-emitting $0\nu\beta\beta$ decay modes that are more stringent than results of previous EXO-200~\cite{Albert:2014fya} and KamLAND-Zen~\cite{Gando:2012pj} searches.

\begin{acknowledgments}
EXO-200 is supported by the DOE and NSF in the U.S., the NSERC in Canada, the SNF in Switzerland, the IBS in Korea, the RFBR in Russia, the DFG in Germany, and CAS and ISTCP in China. The EXO-200 data analysis and simulation use resources of the National Energy Research Scientific Computing Center (NERSC). We gratefully acknowledge the KARMEN collaboration for supplying the cosmic-ray veto detectors, as well as the WIPP for their hospitality.
We would also like to thank Ricardo Cepedello for providing some of the energy spectra of two electrons for neutrinoless double-beta decay of $^{136}$Xe with nonstandard Majoron emission. 
\end{acknowledgments}

\bibliographystyle{apsrev4-1}
\bibliography{reference}% Produces the bibliography via BibTeX.

\end{document}